\documentclass[12pt,a4paper]{article}
\usepackage[margin=27mm]{geometry}
\usepackage[T1]{fontenc}
\usepackage[utf8]{inputenc}
\usepackage{lmodern,microtype,amsmath,amssymb,amsthm,mathtools,booktabs,enumitem}
\usepackage[authoryear,round]{natbib}
\usepackage[hidelinks,breaklinks]{hyperref}
\hypersetup{pdftitle={Endogenous Vindication: Reputation and Effort in Expert Advice},pdfauthor={Georgy Lukyanov and Anna Vlasova},pdfsubject={Analytical working paper},pdfkeywords={expert advice, reputation, effort, career concerns, learning}}
\usepackage{url,needspace}
\setlist[enumerate]{label=(\roman*),leftmargin=2em,itemsep=3pt,topsep=4pt}
\newtheorem{theorem}{Theorem}[section]
\newtheorem{proposition}[theorem]{Proposition}
\newtheorem{lemma}[theorem]{Lemma}
\newtheorem{corollary}[theorem]{Corollary}
\theoremstyle{definition}
\newtheorem{assumption}[theorem]{Assumption}

\newcommand{\E}{\mathbb E}
\newcommand{\Pp}{\mathbb P}
\newcommand{\M}{\mathcal M}
\newcommand{\Q}{\mathcal Q}
\newcommand{\K}{\mathcal K}
\newcommand{\Id}{\mathcal I}
\newcommand{\piI}{\underline\pi}
\newcommand{\Sset}{\mathcal S}
\newcommand{\norm}[1]{\left\lVert #1\right\rVert_\infty}
\newcommand{\KL}{D_{\mathrm{KL}}}

\numberwithin{equation}{section}
\title{Endogenous Vindication:\\Reputation and Effort in Expert Advice\thanks{We are grateful to Costas Cavounidis, Olivier Gossner, Johannes H{\"o}rner, Margarita Kirneva, Yukio Koriyama, Chiara Margaria, Harry Pei, Fran\c{c}ois Salani\'e, Ludvig Sinander, Francesco Squintani, Bruno Strulovici, Stepan Svistunov, Nikhil Vellodi, Maria Ziskelevich, and participants at the 2021 SET seminar and the TSE Micro Theory Workshop for helpful comments and discussions. Financial support from the French National Research Agency (ANR) under grant ANR-17-EURE-0010 (Investissements d'Avenir program) is gratefully acknowledged. All remaining errors are our own.}}
\author{Georgy Lukyanov\thanks{Toulouse School of Economics, 1 Esplanade de l'Universit\'e, 31080 Toulouse Cedex 06, France. Email: \href{mailto:georgy.lukyanov@tse-fr.eu}{georgy.lukyanov@tse-fr.eu}. Corresponding author.}
\and Anna Vlasova\thanks{CREST--\'Ecole Polytechnique, 5 Avenue Henry Le Chatelier, 91120 Palaiseau, France. Email: \href{mailto:anna.vlasova@polytechnique.edu}{anna.vlasova@polytechnique.edu}.}}
\date{}
\begin{document}
\maketitle
\begin{abstract}
Does making advice cheaper to implement increase the supply of advice? We study an expert who can release a favorable certificate to a sequence of clients. Implementation requires the certificate, and its outcome depends on client effort. The expert values client surplus and career income from routine employment; ability is unknown to both the expert and the public. A lower fixed implementation cost weakly increases both advice and client surplus. A lower marginal effort cost can have the opposite effect: greater effort makes failure more revealing and can induce the expert to withhold advice permanently. We establish this reversal for every discount factor on an open family with smooth income utility with constant absolute risk aversion. At a common initial reputation, cheaper effort then eliminates public learning and client surplus. The reversal is impossible when the effort rule is held fixed across the same cost change. Rare outside reports preserve the discounted client-surplus reversal, although recurrent informative reports restore eventual learning. Additional shape restrictions yield a reputation cutoff; without them, advice can be disconnected and even success can stop further evaluation.
\end{abstract}
\clearpage

\section{Introduction}\label{sec:introduction}
An expert's recommendation is often needed before a project can proceed. An engineer may have to certify a proposed change, or an internal specialist may have to approve a trial before a client undertakes it. The outcome then depends on how much effort the client puts into implementation. More effort makes success more likely, but it also changes the interpretation of failure: a failed project is harder to attribute to poor execution. Does making implementation cheaper necessarily encourage the expert to recommend more projects?

This paper studies the distinction between two ways of reducing implementation costs. A lower fixed setup cost raises the value of a project without changing how it is carried out. A lower marginal effort cost also changes the client's behavior. We show that these interventions can have opposite effects on the supply of advice. Lowering the fixed cost weakly increases both advice and expected client surplus. Lowering the marginal cost can instead eliminate advice, even though every project would be more productive if it were undertaken.

We consider a long-lived expert who advises a sequence of short-lived clients. The expert's ability is fixed but unknown to both the expert and the public. Each project has a favorable or an unfavorable state. The expert receives a binary assessment whose accuracy depends on ability. A favorable assessment provides a verifiable certificate, which the expert can release or withhold. Implementation requires this certificate. A client who implements chooses effort, and success requires both a favorable state and sufficient effort. Project outcomes are public and affect the expert's reputation.

The expert values client surplus and also earns career income from routine work. Competitive spot employers pay the expected product of this work, making the wage affine in reputation. Concave income utility makes uncertainty about future wages costly. Routine work continues when a discretionary project is withheld, so silence preserves a valuable income stream. This foundation gives economic content to the career payoff, while retaining explicit restrictions on contracts, insurance, and the information generated by routine work.

The reputational cost of evaluation is familiar. \citet{holmstrom1986} show how concern for human capital can discourage investments that reveal managerial ability. Our contribution concerns a change in the cost of implementing someone else's advice. We first hold the effort rule fixed across a cost change. In that benchmark, a reduction in cost cannot remove a previously optimal recommendation. We then let the client optimize effort. With quadratic effort costs and income utility with constant absolute risk aversion, we construct an open family in which the lower-cost economy remains silent forever at a common initial reputation, while the higher-cost economy supplies advice throughout an upper range of reputations. The construction works for every discount factor and accounts for the full continuation value. It requires neither a cutoff policy nor a decreasing reputational cost of evaluation.

The mechanism follows from how effort changes the two possible outcomes. A success has the same likelihood ratio under every positive effort level. A failure is more informative when the client works harder. Cheaper effort therefore raises current project surplus and lowers the reputation that follows failure. When career income is sufficiently important and income utility sufficiently curved, the second effect dominates both the surplus gain and the value of future opportunities. The expert withholds the certificate, and the client cannot undertake the project. The result establishes a possible reversal on an open parameter family; it does not say that marginal-cost reductions usually discourage advice.

The comparison has direct implications for learning and client welfare. At the initial reputation used in the construction, the lower-cost economy produces no information and no project surplus. The higher-cost economy produces information immediately, gives a high-ability expert a positive chance of eventual vindication, and yields strictly positive expected discounted client surplus. These welfare statements concern the clients' net project returns. They do not include the expert's utility of income or the fiscal cost of a subsidy.

We also separate persistent withholding from permanent failure to learn. The baseline assumes that routine work generates no public information. If its output is instead reported with a small positive probability, the initial advice choices and the client-surplus reversal survive. However, recurrent informative reports eventually reveal ability in both cost economies. The economic cost of delayed advice can therefore remain even when permanent failure of vindication disappears.

Finally, we identify which policy and learning results require additional assumptions. Stronger career concerns shrink the advice set without a cutoff restriction. A cutoff follows when the posterior kernel is ordered and the reputational cost of evaluation decreases over implementable reputations. We give sufficient primitive conditions for this case, but also construct a smooth strictly concave example with disconnected advice. In that example, a success can move the expert into a region where further advice is withheld, and a lower active region can have no path to eventual vindication. These results explain why a cutoff cannot be inferred from concavity or from the informativeness of effort alone.

\subsection{Related literature}\label{sec:literature}
The closest antecedents concern career risk and the choice to expose advice to evaluation. In \citet{holmstrom1986}, investment affects the information used to assess managerial ability, so career concerns can distort project choice. We use this career-risk force rather than claim a new source of aversion to evaluation. \citet{camara2023} study an expert who can avoid judgment by recommending inaction. Consumer participation determines whether feedback becomes available, and their analysis also compares monitoring precision. Thus, neither evaluation avoidance nor endogenous feedback is by itself a distinguishing feature of the present model. Their message game and affine payoff in the ability posterior differ from our certification setting and concave income payoff. Our comparison changes a client cost primitive while holding career preferences fixed, and establishes the opposite cost order when the effort response is removed.

There are also close connections to work on feedback and organizational decisions. \citet{liusanyal2012} show that a principal's decision to cancel a project can prevent verification of expert advice and that better interim information can reduce welfare. \citet{halackremer2020} study dynamic experimentation with career concerns, including delayed stopping and the effects of public information about project quality. \citet{hansen2013} studies performance feedback and career-motivated effort. These papers establish that the incentives to generate or use information can alter its welfare effects. Here the relevant intervention is a change in the client's real cost of productive effort, with the expert deciding whether implementation can occur at all.

An earlier version of this paper circulated as \emph{Dynamic Delegation with Reputation Feedback} \citep{lukyanovvlasova2025}. That version studied receiver effort and reputation feedback and discussed heterogeneous implementation costs. The present version reconstructs the analysis around certified implementation, common uncertainty about ability, and the comparison between fixed and marginal costs. The feedback between effort and inference is retained from that earlier work; the cost-intervention theorem and its fixed-effort counterfactual are the focus of this revision.

\citet{chenschneemeier2026} identify a related conflict between producing information and using it. More precise disclosure attracts informed trading, but makes a manager more reluctant to follow feedback that reveals her to be uninformed. Their manager privately knows her information quality and values the post-investment stock price. Our expert shares the public's uncertainty about ability and faces risk in future career income. The main comparison here changes the cost of productive client effort: cheaper implementation can prevent the project and its evaluation altogether.\footnote{A different cost comparison follows from their equations (20) and (A.50). Lower information-acquisition costs weakly raise informed entry at each disclosure precision without changing conditional feedback use. Since the value of that feedback is nonnegative, expected firm value weakly increases, including after disclosure is optimized. This is a firm-value comparison; their criterion does not deduct investors' information costs.}

Section~\ref{sec:model} describes the environment. Section~\ref{sec:general} characterizes the expert's disclosure problem. Section~\ref{sec:costs} compares fixed and marginal implementation costs, and Section~\ref{sec:welfare} gives the implications for learning and client surplus. Section~\ref{sec:reports} introduces outside reports. Section~\ref{sec:shape} discusses cutoff policies and their limits. The appendices contain the equilibrium construction, supporting results, and remaining proofs.

\section{The model}\label{sec:model}
\subsection{Projects, assessments, and certified implementation}
Time is discrete and infinite. Ability $\theta\in\{L,H\}$ is fixed and unknown to the expert and the public. Let $\pi_t$ denote their belief before the current assessment, with initial prior $\pi_0\in(0,1)$. Project states $\omega_t\in\{0,1\}$ are independently drawn with probability $1/2$ for each state, independently of ability. Assessments are conditionally independent across projects and satisfy
\begin{equation}\label{eq:symmetry}
\Pp(s_t=1\mid\omega_t=1,\theta)
=\Pp(s_t=0\mid\omega_t=0,\theta)=q_\theta,
\qquad \tfrac12<q_L<q_H<1.
\end{equation}
Consequently, either assessment has probability $1/2$ under either ability. Better experts are more often correct; favorable assessments are not more frequent.\footnote{The balanced state prior and conditional symmetry in \eqref{eq:symmetry} do substantive work. With an unequal state prior, the frequency of a favorable assessment generally depends on ability. An unreleased assessment could then give the expert a private ability belief, so the common-belief reduction would have to be reconsidered.}

A favorable assessment provides a verifiable certificate. After $s_t=1$, the expert can release it, denoted $R$, or withhold it, denoted $S$. After $s_t=0$, only $S$ is feasible. A project can be implemented only when its favorable certificate is released: $x_t=0$ is the only feasible action after $S$. After $R$, the short-lived receiver chooses $x_t\in\{0,1\}$ and, if implementing, effort $e_t\in[0,1]$. Implementation and its outcome are public; effort is not. Success $y_t=1$ has probability $\omega_t e_t$. Unimplemented projects expire without an observed outcome. Neither the expert nor the public separately observes the project state, the client's private effort, or the client's net payoff. Apart from the expert's assessment, the information specified here and the routine-work observations described below exhaust observations about ability.

We refer to each short-lived receiver as a client. The client obtains $x_t[y_t-\kappa-c(e_t)]$, where $c\in C^2[0,1]$, $c(0)=c'(0)=0$, $c''>0$, and $c'(1)>q_H$. Write
\begin{equation}\label{eq:receiver}
d=q_H-q_L,\quad p(\pi)=q_L+d\pi,\quad
c'(e(\pi))=p(\pi),\quad W(p)=\max_{e\in[0,1]}\{pe-c(e)\}.
\end{equation}
Assume $W(q_L)<\kappa<W(q_H)$. There is a unique $\piI\in(0,1)$ such that $W(p(\piI))=\kappa$. The receiver implements after $R$ exactly on $I=[\piI,1]$, selecting implementation at equality. Define
\begin{equation}\label{eq:g}
g(\pi)=W(p(\pi))-\kappa\quad(\pi\in[0,1]),
\qquad G=g(1)>0.
\end{equation}
The envelope theorem gives $g'(\pi)=d e(\pi)>0$.

\paragraph{Scope of certification.}
Both the inability to fabricate a favorable certificate and its necessity for implementation are substantive restrictions. They describe an approval process in which a project cannot proceed without favorable documentation. They exclude recommendations after unfavorable assessments and hence exclude excessive testing through fabricated advice. If a client could implement after $S$, the expert would know whether $S$ concealed a favorable or an unfavorable assessment; its outcome could then create a private belief about ability. The one-belief analysis below does not cover that larger game.

\subsection{Routine work and career income}\label{sec:routine}
The expert also performs one unit of routine service every period. It does not compete with the discretionary project for capacity, and its effort is fixed and observable. Fresh risk-neutral spot employers compete for the service. Output $Z\in\{0,z\}$ satisfies $\Pp(Z=z\mid\theta)=a_\theta$, with $0<a_L<a_H<1$. It is conditionally independent of project outcomes and of other routine outputs, given ability.

Employers pay a flat wage before current outcomes. Competition within this spot-contract market gives
\begin{equation}\label{eq:wage}
w(\pi)=w_L+\Delta w\pi,\qquad w_\theta=za_\theta,
\qquad \Delta w=w_H-w_L>0.
\end{equation}
Employment dominates the expert's fixed outside option even at $w_L$. The outgoing employer privately observes routine quality and exits. Future employers and the expert receive no additional private observation of that quality. At baseline, no routine outcome enters the public record. Quality is not contractible in the spot wage, and saving, borrowing, and long-term career insurance are unavailable. These restrictions sustain the one-state model; they are not derived from competition.

Let $U'>0$ and $U''\leq0$ on the entire wage interval. After subtracting an irrelevant constant, the expert evaluates
\begin{equation}\label{eq:utility}
\E\sum_{t\geq0}\delta^t\left\{
\beta r(\pi_t)+\alpha x_t[y_t-\kappa-c(e_t)]\right\},
\quad
r(\pi)=U(w_L+\Delta w\pi)-U(w_L),
\end{equation}
where $\alpha>0$, $\beta\geq0$, and $\delta\in(0,1)$. The coefficient $\alpha$ measures professional concern for another person's surplus, not a component of the expert's own income.\footnote{If the expert instead received a project fee entering income utility $U$, implementation would change current income risk as well as future career income. The additive concern for client surplus in \eqref{eq:utility} keeps those channels separate.} This preference does not provide a separate private signal of the client's realized net payoff. Competition makes wages affine in beliefs; income risk aversion supplies their concave utility. It does not arise from competition alone.

Routine wages are paid before the assessment and discretionary project. A current project therefore affects subsequent career income. The baseline has no public learning while projects are withheld. Section~\ref{sec:reports} relaxes this information restriction.

\subsection{Common beliefs and feasible deviations}\label{sec:beliefs}
Let a public history record messages, implementation, and implemented outcomes. Wages are a known function of the public belief. A certificate reveals $s=1$, but does not by itself reveal ability. At any current common belief $\pi$, an implemented certificate generates posterior maps
\begin{align}
\phi^+(\pi)&=\frac{\pi q_H}{p(\pi)},\label{eq:plus}\\
\phi^-(\pi)&=\frac{\pi[1-e(\pi)q_H]}{1-e(\pi)p(\pi)}.\label{eq:minus}
\end{align}
Put $P(\pi)=e(\pi)p(\pi)$ and define, on the full belief interval,
\begin{equation}\label{eq:M}
\M f(\pi)=P(\pi)f(\phi^+(\pi))+
[1-P(\pi)]f(\phi^-(\pi)).
\end{equation}
Both endpoints are fixed and $\M\iota(\pi)=\pi$ for $\iota(\pi)=\pi$. For an interior prior, $\phi^-<\pi<\phi^+$.

\begin{proposition}\label{prop:foundation}
The game admits a sequentially rational assessment with the following properties. Expert and public ability beliefs agree after every feasible public history. An unexpected feasible certificate leaves that ability belief unchanged before its outcome. The receiver follows \eqref{eq:receiver}, and the expert has a stationary optimal disclosure rule. The expert's bounded continuation value is unique. Beliefs in this assessment are limits of Bayes beliefs from completely mixed feasible message and implementation choices and full-support effort perturbations.
\end{proposition}
\noindent The proof is in Appendix~\ref{app:foundations}.

The proposition establishes a consistent assessment and a unique value for the associated decision problem. It does not assert uniqueness of all perfect Bayesian assessments with unrestricted off-path beliefs. Indifference can also allow more than one expert strategy. Throughout, the expert releases an eligible certificate at indifference and withholds an unimplementable certificate.

On $I$, the expert's Bellman equation is
\begin{equation}\label{eq:bellman}
V(\pi)=\beta r(\pi)+\frac\delta2 V(\pi)
+\frac12\max\{\delta V(\pi),\alpha g(\pi)+\delta\M V(\pi)\}.
\end{equation}

\section{Advice and career risk}\label{sec:general}
Let $r$ be bounded, continuous, and increasing on $[0,1]$. The wage foundation gives a concave payoff; we will also consider affine and convex payoffs to identify the boundary of the mechanism. Define
\begin{equation}\label{eq:J}
J_r(\pi)=r(\pi)-\M r(\pi),\qquad
\lambda=\frac\beta\alpha,\qquad
\chi_\lambda=\frac{\delta\lambda}{1-\delta}.
\end{equation}
Subtract the value $B(\pi)=\beta r(\pi)/(1-\delta)$ of permanent withholding and write $v_\lambda=(V-B)/\alpha$. Permanent withholding is feasible, so $v_\lambda\geq0$.

\begin{proposition}\label{prop:option}
The normalized option value is the unique bounded fixed point of
\begin{equation}\label{eq:T}
\mathcal T_\lambda f(\pi)=
\begin{cases}
0,&\pi<\piI,\\[2pt]
\displaystyle\max\left\{0,
\frac{g(\pi)-\chi_\lambda J_r(\pi)+\delta\M f(\pi)}{2-\delta}\right\},&\pi\in I.
\end{cases}
\end{equation}
The operator is order preserving and contracts by $\delta/(2-\delta)$. Let
\begin{equation}\label{eq:A}
A_\lambda=g-\chi_\lambda J_r+\delta\M v_\lambda\quad\text{on }I.
\end{equation}
The stationary advice set is $\Sset_\lambda=\{\pi\in I:A_\lambda(\pi)\geq0\}$. The actual normalized gain from release rather than waiting is
\begin{equation}\label{eq:D}
D_\lambda=\frac{\alpha g+\delta\M V-\delta V}{\alpha}
=\begin{cases}
A_\lambda,& A_\lambda\leq0,\\[2pt]
\displaystyle\frac{2(1-\delta)}{2-\delta}A_\lambda,&A_\lambda>0.
\end{cases}
\end{equation}
If $J_r\geq0$ on $I$, then
\begin{equation}\label{eq:bound}
0\leq v_\lambda\leq\frac{G}{2(1-\delta)}.
\end{equation}
\end{proposition}
\noindent The proof is in Appendix~\ref{app:foundations}.

All evaluations of $\M$ use the full posterior interval. In particular, $v_\lambda(\phi^-)=0$ when failure takes the belief below $\piI$. Concavity of $r$ merely on $I$ does not suffice for Jensen's inequality at such a prior. Concavity on $[0,1]$ is a transparent sufficient condition for $J_r\geq0$; the latter inequality on $I$ is the weaker condition used below.

\begin{corollary}\label{cor:boundary}
If $\lambda=0$, or if $J_r\leq0$ throughout $I$, every implementable certificate is strictly worth releasing, including at $\piI$. In particular, affine or convex $r$ does not produce withholding in this stationary baseline. More generally, if all net rewards are bounded nonnegative rewards $k(\pi)$ earned only from implemented advice, with no other flow or learning during silence, release is strict wherever $k(\pi)>0$.
\end{corollary}
\noindent The proof is in Appendix~\ref{app:foundations}.

The reward-only benchmark uses stationarity, an unchanged state during silence, and nonnegative incremental rewards. It does not cover finite horizons, negative current rewards, changing opportunities, wealth accumulation, or background information. Endogenous continuation value alone does not create the withholding mechanism in this benchmark.

The index in \eqref{eq:A} separates the three forces behind a recommendation. The client gains the current surplus $g$. Evaluation exposes future career income to the Jensen loss $J_r$, weighted by $\chi_\lambda$. Finally, the new posterior may create valuable opportunities for future advice, represented by $\delta\M v_\lambda$. Concavity makes the second term costly, but it does not by itself make the index increasing in reputation.

The bound in \eqref{eq:bound} also gives
\begin{equation}\label{eq:totalbound}
g+\delta\M v_\lambda\leq \frac{G(2-\delta)}{2(1-\delta)}.
\end{equation}
This upper bound includes all future advice opportunities. We use it below to compare implementation costs without solving for the value function explicitly. For strictly concave $r$, sufficiently strong career concerns suppress advice on any compact interval of interior implementable beliefs, while advice remains strict near certainty. Proposition~\ref{prop:suppression} states these observations precisely.

\section{Implementation costs and the supply of advice}\label{sec:costs}
\subsection{Holding the experiment fixed}
For a fixed career weight, write $v=v_\lambda$, $\chi=\chi_\lambda$, $J=J_r$, and $A=A_\lambda$. Subscripts distinguish cost economies. We first compare rewards while keeping the outcome experiment unchanged.

\begin{proposition}\label{cost:prop:order}
Fix $r,\chi,\delta$ and the posterior kernel $\M$. Suppose two bounded Borel project-reward functions satisfy $g_2\leq g_1$, with eligibility $I_i=\{g_i\geq0\}$. Then their stationary release sets satisfy $\Sset_2\subseteq\Sset_1$. No decreasing-Jensen-cost or cutoff assumption is needed. Their public records are ordered by a type-independent censoring: the record under $\Sset_2$ is a garbling of the record under $\Sset_1$.
\end{proposition}
\begin{proof}
Let $T_i$ be the operator in \eqref{eq:T}. On $I_2$, the two operators differ only in $g_i$, so $T_2f\leq T_1f$. On $I_1\setminus I_2$, $T_2f=0\leq T_1f$; outside $I_1$ both are zero. Monotone iteration from zero gives $v_2\leq v_1$. Therefore $A_2\leq A_1$ on $I_2$, proving nesting with common ties. Couple signals and outcomes until the larger-set record first reaches a belief outside $\Sset_2$. The smaller-set process then stays silent forever; replace all subsequent observations by silence. This censoring depends only on the public history, not on ability, and produces exactly the smaller-set law. It also covers an initially inactive smaller-set prior. The sign of $J$ is not needed for the operator comparison.
\end{proof}
\begin{corollary}\label{cost:cor:fixed}
\leavevmode
\begin{enumerate}
\item With optimal effort and unchanged $c$, reducing $\kappa$ weakly expands advice and makes public records weakly more informative. Expected discounted client surplus weakly increases.
\item Fix any feasible Borel effort profile $\bar e(\pi)$ across two economies with costs $c_i(e)=\zeta_i e^2/2$ and $\zeta_2>\zeta_1$. Holding all other primitives fixed, the higher-cost economy has a weakly smaller release set. A marginal-cost reduction cannot eliminate previously optimal release in this benchmark.
\end{enumerate}
\end{corollary}
\begin{proof}
A setup-cost change leaves the optimal effort and experiment unchanged and changes $g$ by a constant. Proposition \ref{cost:prop:order} applies. Couple the two policies until the higher-setup-cost policy stops. Each common project has larger expected net surplus under the lower setup cost, and any additional projects have nonnegative conditional expected surplus. Boundedness and discounting justify summation. For the second assertion,
\[
 g_i^F(\pi)=p(\pi)\bar e(\pi)-\frac{\zeta_i\bar e(\pi)^2}{2}-\kappa.
\]
Thus $g_2^F\leq g_1^F$, while $\M$ remains fixed. Apply Proposition \ref{cost:prop:order} again.
\end{proof}
The frozen profile may be constant or depend on reputation, including a profile equal to optimal effort in one of the two economies. This is a constrained benchmark: implementation remains a choice, but the effort rule is imposed and need not be optimal in both economies. It holds the experiment fixed under the same real cost intervention.\footnote{An exogenous kernel deliberately indexed by cost could reproduce an endogenous-effort reduced form. The comparison rules out the reversal when the effort rule and hence the experiment are unchanged across the cost intervention; it does not establish an impossibility result for every exogenous-precision model.}

\subsection{The response of effort and information}
\begin{lemma}\label{lem:effort}
At a fixed interior prior, raising implemented effort makes the binary outcome experiment more informative in the Blackwell order. Its two ability-specific Kullback--Leibler information rates strictly increase. If $r$ is concave on the full posterior support, $J_r(\pi;e)$ weakly increases with effort. For $r(\pi)=A+B\pi-\Gamma\pi^2$, $\Gamma>0$,
\begin{equation}\label{eq:effortJ}
J_r(\pi;e)=\Gamma\frac{d^2\pi^2(1-\pi)^2e}{p(\pi)[1-ep(\pi)]},
\qquad
\frac{\partial J_r}{\partial e}=
\Gamma\frac{d^2\pi^2(1-\pi)^2}{p(\pi)[1-ep(\pi)]^2}>0.
\end{equation}
\end{lemma}
\begin{proof}
From the experiment at $e_2>e_1>0$, retain a success with probability $e_1/e_2$, otherwise report failure; retain every original failure. Under either ability this produces the experiment at $e_1$. A garbled failure pools two positive-probability outcomes with distinct likelihood ratios, so the data-processing inequality for relative entropy is strict in either direction. The posterior from the garbled experiment is the conditional expectation of the original posterior; conditional Jensen gives the career-cost comparison. Finally,
\[
\operatorname{Var}(\pi'\mid\pi;e)
=\frac{d^2\pi^2(1-\pi)^2e}{p(\pi)[1-ep(\pi)]},
\]
and a quadratic payoff's Jensen cost is $\Gamma$ times this variance. Differentiation gives \eqref{eq:effortJ}.
\end{proof}

The comparison holds at a fixed prior. It does not imply that $J_r(\pi;e(\pi))$ decreases with reputation. The direct effort effect increases exposure risk. A decreasing total Jensen cost requires the concentration of beliefs near certainty to dominate that effect.

\subsection{Cheaper effort can eliminate advice}
For $c_i(e)=\zeta_i e^2/2$, where $\zeta_i>q_H$, optimal effort and surplus are
\[
 e_i(\pi)=\frac{p(\pi)}{\zeta_i},\qquad
 g_i(\pi)=\frac{p(\pi)^2}{2\zeta_i}-\kappa.
\]
Subscripts 1 and 2 denote cheaper and costlier effort, respectively. Choose
\begin{equation}\label{cost:eq:primitive}
 q_H<\zeta_1<\zeta_2,\quad
 \frac{\zeta_2}{\zeta_1}<\left(\frac{q_H}{q_L}\right)^2,\quad
 \frac{q_L^2}{2\zeta_1}<\kappa<\frac{q_H^2}{2\zeta_2}.
\end{equation}
These strict conditions are feasible: take $\zeta_2$ sufficiently close to any $\zeta_1>q_H$. Both eligibility thresholds are interior, with $\pi_{I,1}<\pi_{I,2}$. Put $G_i=q_H^2/(2\zeta_i)-\kappa$.

\begin{lemma}\label{cost:lem:failure}
At every interior prior, $\phi^-_1(\pi)<\phi^-_2(\pi)$. For either quadratic cost, $\phi^-_i$ is strictly increasing in the prior.
\end{lemma}
\begin{proof}
At a fixed prior,
\[
 \frac{\partial\phi^-}{\partial e}
 =-\frac{d\pi(1-\pi)}{(1-ep)^2}<0.
\]
Since $e_1>e_2$, the first claim follows. For the second, posterior odds equal
\[
 \frac{\phi^-_i}{1-\phi^-_i}
 =\frac{\pi}{1-\pi}\frac{\zeta_i-pq_H}{\zeta_i-pq_L}.
\]
The derivative of their logarithm is
\[
 \frac{1}{\pi(1-\pi)}-
 \frac{d^2\zeta_i}{(\zeta_i-pq_H)(\zeta_i-pq_L)}>0.
\]
To see the inequality, $\zeta_i-pq_H>q_H-p$ and
$1-pq_L/\zeta_i>p-q_L$, using $\zeta_i>q_H$, $p<q_H<1$.
Their product exceeds $(q_H-p)(p-q_L)=d^2\pi(1-\pi)$. Endpoint values follow by continuity.
\end{proof}

\begin{theorem}\label{cost:thm:reversal}
Fix any $\delta\in(0,1)$ and cost primitives satisfying \eqref{cost:eq:primitive}. For any $\pi_0\in(\pi_{I,2},1)$, there exist finite $\rho>0$, a common career weight $\chi>0$, and $b\in(\pi_{I,2},\pi_0)$ such that, for the same smooth income payoff
\[
 r_\rho(\pi)=1-\exp(-\rho\pi),
\]
economy 1 strictly withholds at $\pi_0$ forever, whereas economy 2 strictly releases every favorable certificate at all $\pi\in[b,1]$. The inequalities hold on an open set of the finite-dimensional primitives and of initial priors within the maintained symmetric-assessment model. In particular, endogenous effort reverses the frozen-effort cost order in Corollary \ref{cost:cor:fixed}.
\end{theorem}
\begin{proof}
Write $f_1=\phi^-_1(\pi_0)$ and $s_1=1-P_1(\pi_0)>0$. Lemma \ref{cost:lem:failure} and continuity allow $b<\pi_0$, still above $\pi_{I,2}$, sufficiently close to $\pi_0$ that $f_2:=\phi^-_2(b)>f_1$. Also $f_1<\pi_0$.
For exponential utility,
\[
 J_i(\pi)=P_i e^{-\rho\phi^+}+(1-P_i)e^{-\rho\phi^-_i}-e^{-\rho\pi}.
\]
At $\pi_0$ in economy 1,
\[
 J_1(\pi_0)\geq e^{-\rho f_1}
 \left[s_1-e^{-\rho(\pi_0-f_1)}\right]
 \geq\frac{s_1}{2}e^{-\rho f_1}
\]
whenever $\rho>\log(2/s_1)/(\pi_0-f_1)$. In economy 2, all posteriors from $\pi\geq b$ are at least $f_2$, so concavity and the formula give the uniform bound
\[
 0\leq J_2(\pi)\leq e^{-\rho f_2}\quad(\pi\geq b).
\]
Let $L_1=G_1(2-\delta)/[2(1-\delta)]$. Choose $\rho$ large enough also to satisfy
\[
 e^{\rho(f_2-f_1)}>\frac{2L_1}{s_1g_2(b)}.
\]
Both requirements are possible because $f_2-f_1>0$ and $g_2(b)>0$. The following open interval is therefore nonempty; choose the common $\chi$ in it:
\begin{equation}\label{cost:eq:chi}
 \frac{2L_1}{s_1}e^{\rho f_1}<\chi<g_2(b)e^{\rho f_2}.
\end{equation}
The upper continuation bound \eqref{eq:totalbound} yields
\[
 A_1(\pi_0)\leq L_1-\chi\frac{s_1}{2}e^{-\rho f_1}<0.
\]
Conversely, $v_2\geq0$ and increasing $g_2$ yield
\[
 A_2(\pi)\geq g_2(b)-\chi e^{-\rho f_2}>0\quad(\pi\geq b).
\]
Silence leaves the prior unchanged, so the first strict choice repeats forever. Set $\beta/\alpha=\chi(1-\delta)/\delta$ in both economies to implement the same career concern. All displayed sufficient inequalities are strict and continuous in their finite parameters, including $\pi_0$, with $b$ retained as a witness in a sufficiently small neighborhood. They therefore define an open family. This last argument uses the explicit bounds, not an unproved continuity claim for a possibly discontinuous optimal policy.
\end{proof}

The two cost economies share the same $\rho$ and $\chi$. The quantifier over discount factors is pointwise.\footnote{The parameters $\rho$ and $\chi$ may depend on $\delta$, as well as on the cost primitives and initial prior. The theorem does not require one fixed parameter choice to work uniformly as $\delta$ approaches zero or one.}

\noindent\textbf{Income foundation.} With routine wage $w(\pi)=w_L+\Delta w\pi$, $\Delta w>0$, let $U(w)=1-\exp[-\eta(w-w_L)]$ and $\eta=\rho/\Delta w$. Then $U(w(\pi))-U(w_L)=r_\rho(\pi)$ and $-U''/U'=\eta$ is constant. The same income utility and career weight apply to both economies. The construction allows large risk aversion and career weight; the theorem establishes possibility on an open family rather than a claim about empirical prevalence.

\noindent\textbf{Economic mechanism.} Cheaper effort raises both current surplus and per-project informativeness. In this success technology, the success posterior is independent of effort. More effort instead makes failure a worse signal of ability. Under CARA, the career loss associated with the lower failure posterior can outweigh the surplus improvement and every possible continuation gain. The bound in \eqref{cost:eq:chi} makes this comparison exact. Curvature and career concerns supply the aversion to evaluation. The effort response changes how that aversion reacts to the client's cost. This is the distinction from the fixed-experiment comparison.

\section{Learning and client surplus}\label{sec:welfare}
Let $O(\pi)=\pi/(1-\pi)$ and define ex-ante discounted client welfare by
\[
 \mathcal W_i(\pi_0)=\E_{\pi_0}\sum_{t=0}^{\infty}\delta^t
 x_t\left[y_t-\kappa-\frac{\zeta_i e_t^2}{2}\right].
\]
The expectation uses the common initial prior, rather than conditioning on high ability. Define high-ability vindication as $\pi_t\to1$.
\begin{corollary}\label{cost:cor:welfare}
For the economies in Theorem \ref{cost:thm:reversal}, economy 1 has $\mathcal W_1(\pi_0)=0$ and no high-ability vindication. In economy 2,
\[
 \mathcal W_2(\pi_0)\geq\frac{g_2(\pi_0)}2>0,
 \qquad
 \Pp_H(\pi_t\to1)\geq1-\frac{O(b)}{O(\pi_0)}>0.
\]
Its public record is strictly more informative about ability than the constant record in economy 1, already over the first period. These are comparisons at the constructed common initial prior, not global orderings for all priors or all costs.
\end{corollary}
\begin{proof}
Permanent silence in economy 1 proves the zero statements. In economy 2 a favorable certificate arrives with probability $1/2$ at time zero and yields expected net surplus $g_2(\pi_0)>0$. Every subsequent implemented project has nonnegative conditional expected surplus, which gives the welfare bound.

Until first leaving $[b,1]$, economy 2 always releases favorable certificates. Under $H$, the likelihood ratio $O(\pi_0)/O(\pi_t)$ is a nonnegative martingale. Its maximal inequality bounds the probability of ever crossing below $b$ by $O(b)/O(\pi_0)$. On paths that remain above $b$, informative tests continue, with efforts in a compact interval bounded away from zero. Conditional high-ability log-likelihood increments have a uniformly positive mean (one half the Bernoulli relative entropy for an implemented test) and are bounded. The martingale strong law implies log odds tend to infinity. Hence these paths produce vindication, proving the lower bound. Further active regions below $b$ can only add paths to vindication.

At time zero, public success has probability $e_2(\pi_0)q_\theta/2$, which differs by type. Thus the public record is a nontrivial experiment. Ignoring it produces economy 1's constant record, whereas a constant record cannot generate its type-dependent distribution by type-independent randomization. The comparison persists for every horizon containing that first period.
\end{proof}
This is a client-surplus result, not an aggregate-welfare theorem including the expert's income utility.\footnote{The intervention changes the real cost of effort. Interpreting it as a subsidy would additionally require fiscal accounting and a specification of who pays. The displayed welfare criterion does not include that fiscal cost.} Nor does the result say that larger effort makes a given test less informative: each test is more informative when effort is cheaper; the expert may stop supplying tests altogether.

\section{Outside reports and delayed advice}\label{sec:reports}
The no-learning condition during silence is separate from career-income concavity. Suppose each routine output is publicly reported after the project stage with probability $\varepsilon\in[0,1]$, independently across dates and independently of ability, output, and the project history. An absent report therefore conveys no additional information. Given belief $\pi$, put
\[
a(\pi)=a_L+(a_H-a_L)\pi,\quad
\psi_z(\pi)=\frac{\pi a_H}{a(\pi)},\quad
\psi_0(\pi)=\frac{\pi(1-a_H)}{1-a(\pi)}.
\]
Let $\K f=a f(\psi_z)+(1-a)f(\psi_0)$ and $\Q_\varepsilon=(1-\varepsilon)\Id+\varepsilon\K$. Silence now has transition $\Q_\varepsilon$; implemented advice has transition $\M\Q_\varepsilon$, with the project posterior formed first. The wage remains based on the pre-outcome belief.

On $I$, the new value satisfies
\begin{equation}\label{eq:reportsBellman}
V_\varepsilon=\beta r+\frac\delta2\Q_\varepsilon V_\varepsilon
+\frac12\max\{\delta\Q_\varepsilon V_\varepsilon,
\alpha g+\delta\M\Q_\varepsilon V_\varepsilon\};
\end{equation}
below $I$, $V_\varepsilon=\beta r+\delta\Q_\varepsilon V_\varepsilon$. The unnormalized release gain is
\begin{equation}\label{eq:reportsD}
\mathcal D_\varepsilon=
\alpha g+\delta(\M-\Id)\Q_\varepsilon V_\varepsilon.
\end{equation}
Its baseline value is $\mathcal D_0=\alpha D_\lambda$. The passive value is now $(\Id-\delta\Q_\varepsilon)^{-1}\beta r$, not the pointwise annuity. In particular, the baseline nesting, cutoff, and no-distortion claims are not automatically results for this altered transition system.

\begin{proposition}\leavevmode\label{prop:reports}
Normalize $r(0)=0$ and put $R=\max r$, $C_0=\beta R+\alpha G/2$. Then
\begin{align}
\norm{V_\varepsilon-V_0}&\leq
\frac{\delta\varepsilon C_0}{(1-\delta)^2},\label{eq:reportsVbound}\\
\sup_I|\mathcal D_\varepsilon-\mathcal D_0|&\leq
\frac{2\delta\varepsilon C_0}{(1-\delta)^2}.\label{eq:reportsDbound}
\end{align}
Consequently, any uniform strict action margin persists for sufficiently small $\varepsilon$. For every $\varepsilon>0$ and the maintained interior common prior, however, the public posterior converges to the true ability almost surely. If the initial belief remains in a strict withholding region, and $T_{\rm test}$ is the first implemented-project period indexed from zero, then
\begin{equation}\label{eq:wait}
\Pp_\theta(T_{\rm test}>n)\geq(1-\varepsilon)^n\quad(n\geq0),
\qquad \E_\theta T_{\rm test}\geq\varepsilon^{-1}.
\end{equation}
For sufficiently small positive reporting rates, high ability eventually resumes project testing almost surely.
\end{proposition}
\noindent The proof is in Appendix~\ref{app:reports}.

Strict withholding and arbitrarily long expected delays are compatible with eventual learning. Permanent false distrust, including the barrier outcome of Theorem~\ref{thm:noninterval}, relies on the absence of recurring informative reports. The small-reporting limit is continuous in values and strict actions but need not be continuous in the probability of permanent false distrust.

The reporting extension also preserves the implementation-cost comparison in discounted client surplus.
\begin{proposition}\label{cost:prop:reports}
For the strict economies of Theorem \ref{cost:thm:reversal}, there exists $\bar\varepsilon>0$ such that every $0<\varepsilon<\bar\varepsilon$ preserves silence at $\pi_0$ in economy 1, release at $\pi_0$ in economy 2, and
$\mathcal W_2^\varepsilon(\pi_0)>\mathcal W_1^\varepsilon(\pi_0)$.
Recurring informative reports identify ability in both economies. The baseline contrast in permanent high-ability vindication therefore does not survive positive reporting.
\end{proposition}
\noindent The proof is in Appendix~\ref{app:reports}.

The result separates a welfare loss from a permanent learning failure. With rare reports, a client may wait a long time before the expert is willing to supply advice. Later identification does not recover the discounted surplus lost during that delay. However, any positive rate of recurring informative reports eventually identifies ability, so the baseline comparison in permanent vindication cannot be carried over to this extension.

\section{When does advice follow a cutoff?}\label{sec:shape}
The cost reversal was established without imposing an interval-shaped advice set. This matters because the shape of optimal advice requires restrictions beyond concavity. Three further results clarify the distinction.

First, stronger career concerns weakly reduce advice for a fixed experiment. If $J_r\geq0$ on $I$, increasing $\lambda=\beta/\alpha$ lowers the option value and the index in \eqref{eq:A}. Thus the advice sets are nested, whether or not either set is an interval (Theorem~\ref{thm:nesting}). The record under the smaller set can be obtained by retaining the larger-set record until the smaller policy stops and then replacing later observations by silence. This is a type-independent garbling. The argument depends on a common outcome experiment; it is not a comparison across optimized effort costs.

Second, suppose the posterior kernel is increasing in first-order stochastic dominance and $J_r$ is weakly decreasing over implementable reputations. Then the advice index is increasing and advice follows a cutoff (Theorem~\ref{thm:cutoff}). The cutoff remains at the client's implementation threshold for a positive interval of weak career concerns. It subsequently rises weakly with career weight and approaches one. Proposition~\ref{prop:high} gives sufficient smooth conditions when implementation requires high reputation, and Proposition~\ref{prop:quadratic} gives an explicit quadratic family. These are sufficient restrictions on the implementation region. Greater effort, by itself, does not deliver the decreasing Jensen cost that they require.

Third, without the decreasing-cost condition, a success can stop further advice. Theorem~\ref{thm:noninterval} constructs smooth increasing strictly concave income utility for which a lower active region and an upper active region are separated by a withholding interval. A success can raise reputation into that interval. Moreover, the lower region can be unable to reach the upper one before testing stops. The high-ability expert is then denied eventual vindication with probability one from that lower region, despite initially supplying advice. The posterior kernel remains ordered throughout the construction.

The general learning result in Theorem~\ref{thm:learning} gives the appropriate qualification. A low-ability expert stops testing in finite expected time under every stationary advice set considered here. A high-ability expert is vindicated on paths with continued testing, but such paths need not have positive probability from every active reputation. Positive vindication requires a feasible path to an upper active region through active predecessors. Under a cutoff this path is always available; with disconnected advice it may be blocked. These learning arguments also apply to fixed informative effort. They describe what the supply of tests permits the public to learn, while the cost theorem identifies an intervention that changes that supply through optimized effort.

\section{Discussion and conclusion}\label{sec:discussion}
The model isolates an organizational distinction between paying less to start a project and paying less for the effort that makes it succeed. Lower setup costs leave the interpretation of outcomes unchanged and expand the supply of advice. Lower effort costs improve each implemented project, but also make a failure more damaging to the expert's career. This second effect can be strong enough to prevent implementation altogether. The resulting loss of client surplus survives small amounts of independent public information.

Several restrictions delimit this conclusion. Certification is essential to the action space: a client cannot implement an uncertified project, and the expert cannot fabricate a favorable certificate. The model therefore addresses withholding of favorable assessments, not the possibility of recommending projects after unfavorable assessments. The common state prior and conditional signal symmetry make assessments equally frequent across ability levels. Better experts are more often correct, rather than more often favorable. If the good-state prior were $\mu\ne1/2$, then
\[
\Pp(s=1\mid\theta)=(1-\mu)+(2\mu-1)q_\theta,
\]
so the expert's private assessments would themselves reveal ability. A model with varying signal strength and selective exposure could introduce the same issue. Such extensions require a richer information state and are not covered by the present equilibrium construction.

The career-income foundation is likewise conditional on continued routine employment and limited insurance. Competitive wages alone do not generate concave career utility; income risk aversion does. Allowing the expert to insure future wages, accumulate wealth, or contract jointly over routine service and project outcomes would change the problem. These features may matter in applications, but they are separate from the cost comparison established here.

\appendix
\section{Equilibrium and the option value of advice}\label{app:foundations}

\subsection{Proof of Proposition~\ref{prop:foundation}}
\begin{proof}
Before a new assessment, suppose the expert's private history, conditional on the public history, conveys no further ability information. Equation~\eqref{eq:symmetry} implies that the new assessment is independent of ability and the past. Any message rule based on that private history and the assessment consequently has an ability-independent message probability. Conditional private histories remain uninformative about ability after either message.

After $S$, implementation is infeasible, so no outcome can change this conclusion. After $R$, the assessment is public. The observation restrictions exclude any further private outcome or payoff feedback to the expert. Any receiver action or randomization uses public information only. If implementation occurs, its outcome depends on ability but not on the expert's unrevealed past. Bayesian updating therefore changes expert and public beliefs identically. This proves the induction for all positive-probability histories, including under feasible strategy perturbations. For an unobserved randomized effort, the outcome formulas use its conditional mean. In the equilibrium limit that mean is $e(\pi)$, yielding \eqref{eq:plus}--\eqref{eq:minus}.

For consistency at a zero-probability feasible history, perturb each binary choice toward both feasible actions and each effort choice toward a full-support distribution on $[0,1]$. The perturbations cannot depend on ability, which the players do not observe. The preceding induction applies to every perturbed finite public history. The probability of $R$ or $S$ conditional on either ability is identical, including when its limit is zero. Fix the prescribed strategies at each finite history while taking the perturbation size to zero. The associated mean efforts and Bayesian posterior ratios converge recursively to the stated beliefs. The perturbations establish consistency; they need not be equilibria of the perturbed games.

Given these beliefs, the receiver's strict effort optimum and its implementation rule follow directly from its one-period payoff. Routine wages satisfy the spot-market zero-profit and outbidding conditions. At a belief below $\piI$, no project is implemented, so the expert obtains the career annuity. On $I$, the Bellman equation is \eqref{eq:bellman}. 
This operator, with $V=\beta r+\delta V$ below $I$, is a contraction of modulus $\delta$ on bounded Borel functions. It has a unique bounded Borel fixed point and a measurable optimal selector, since there are only two expert actions. The Bellman upper bound applies after every feasible information set, and iterating it bounds any history-dependent deviation; its bounded terminal remainder vanishes. The stationary selector attains the bound. This verifies sequential rationality and sufficiency of the common belief for the prescribed continuation problem.
\end{proof}

\subsection{Proof of Proposition~\ref{prop:option}}
\begin{proof}
Subtracting $B$ in \eqref{eq:bellman} gives
\[
(2-\delta)v_\lambda
=\max\{\delta v_\lambda,g-\chi_\lambda J_r+\delta\M v_\lambda\}.
\]
If its second argument is negative, the only nonnegative solution at the state is zero. If it is positive, the solution is that argument divided by $2-\delta$; the first argument is then strictly smaller. The equality case also gives zero. This proves \eqref{eq:T}. Conditional expectation and the positive-part operation are nonexpansive, establishing contraction and order preservation. Substitution of $v_\lambda=\max\{0,A_\lambda/(2-\delta)\}$ into $D_\lambda=A_\lambda-\delta v_\lambda$ gives \eqref{eq:D}. If $J_r\geq0$, taking suprema in \eqref{eq:T} gives $(2-\delta)\norm{v_\lambda}\leq G+\delta\norm{v_\lambda}$, proving the bound.
\end{proof}

\subsection{Proof of Corollary~\ref{cor:boundary}}
\begin{proof}
In the first cases $A_\lambda\geq g$, giving strictness above $\piI$. At $\piI$, success has positive probability and $\phi^+(\piI)>\piI$, where $v_\lambda\geq g/(2-\delta)>0$. Thus $A_\lambda(\piI)>0$ as well. For the final claim, permanent withholding yields zero and the reward-only value satisfies
\[
V=\max\{0,(k+\delta\M V)/(2-\delta)\}.
\]
At $k>0$, $V>0$ and the gain from release is $k+\delta\M V-\delta V=2(1-\delta)V>0$.
\end{proof}

\begin{proposition}\label{prop:suppression}
Suppose $r$ is strictly concave on $[0,1]$. For a compact interval $C\subset(\piI,1)$ let $j_C=\min_{\pi\in C}J_r(\pi)>0$. If
\begin{equation}\label{eq:suppression}
\lambda>\frac{G(2-\delta)}{2\delta j_C},
\end{equation}
every certificate at a belief in $C$ is strictly withheld. For every finite $\lambda$, certificates are strictly released sufficiently near one. Neither conclusion requires a decreasing Jensen cost or an ordered kernel.
\end{proposition}
\begin{proof}
Strict concavity and the nondegenerate posterior distribution give $J_r>0$ on $(0,1)$, and continuity gives $j_C>0$. Equations~\eqref{eq:A}--\eqref{eq:bound} imply
\[
A_\lambda(\pi)\leq
\frac{G(2-\delta)/2-\delta\lambda j_C}{1-\delta}<0
\qquad(\pi\in C).
\]
As $\pi\uparrow1$, $g(\pi)\to G>0$ and $J_r(\pi)\to0$. Since $\M v_\lambda\geq0$, the index is positive near one.
\end{proof}

\section{Career concerns and cutoff policies}\label{sec:nesting}
\begin{theorem}\label{thm:nesting}
Fix $g$, $\M$, $\delta$, and $r$. Suppose $J_r\geq0$ on $I$. If $\lambda_2\geq\lambda_1$, then
\[
v_{\lambda_2}\leq v_{\lambda_1},\qquad
A_{\lambda_2}\leq A_{\lambda_1},\qquad
\Sset_{\lambda_2}\subseteq\Sset_{\lambda_1}.
\]
No monotonicity in the prior is required of the Jensen cost, the kernel, the value, or the advice sets.
\end{theorem}
\begin{proof}
For every bounded $f$, $\mathcal T_{\lambda_2}f\leq\mathcal T_{\lambda_1}f$, because $J_r\geq0$. Both operators preserve order. Iteration from zero and uniform convergence therefore order their fixed points. Substitution into \eqref{eq:A} orders the indices. The common weak-inequality tie rule gives the set inclusion.
\end{proof}

\begin{corollary}\label{cor:records}
Consider two stationary advice sets $\Sset_2\subseteq\Sset_1\subseteq I$ with the same implementation rule and outcome kernel. Start from the same prior and let $\tau_i$ be the first date with $\pi_t\notin\Sset_i$. There is a coupling under either ability such that
\[
\tau_2\leq\tau_1,\qquad
\{\tau_2=\infty\}\subseteq\{\tau_1=\infty\}.
\]
For every finite horizon, and for the entire public record, the record under $\Sset_2$ is a garbling of the record under $\Sset_1$. Accordingly, stronger career concerns in Theorem~\ref{thm:nesting} weakly increase cessation under high ability and weakly reduce the information generated about ability.
\end{corollary}
\begin{proof}
Use the same assessments and conditional outcome randomizations until the common belief first leaves $\Sset_2$. Before that time both policies release exactly the same favorable certificates, induce the same effort, and produce the same record. Once the smaller policy stops, its belief is constant and it sends $S$ forever. This proves the pathwise assertions, including when the initial belief is already outside one set.

From a record generated by $\Sset_1$, compute the common Bayesian belief recursively and retain the record up to the first exit from $\Sset_2$. Replace every subsequent message by $S$ and every subsequent project by no implementation and no outcome. This deterministic map uses no knowledge of ability and has exactly the law of the smaller-policy record under either ability. Any reported wage path is recomputed from the censored beliefs. The same map truncated at any finite date proves the finite-horizon claim.
\end{proof}

This is a comparison holding the experiment fixed. An effort-cost change alters both $g$ and $\M$, so it cannot be evaluated by this censoring argument. Also, weak nesting does not imply that advice changes for every increase in career weight.

\subsection{A cutoff under additional restrictions}\label{sec:cutoff}
\begin{assumption}\label{ass:shape}
On $I$, the posterior kernel is increasing in first-order stochastic dominance and $J_r$ is weakly decreasing. The career payoff is continuous on $[0,1]$ and strictly concave on $[0,1]$.
\end{assumption}
The full-support concavity condition is stronger than needed for the order proof, but ensures strict positive Jensen cost at every interior prior, including when failure leaves $I$.

\begin{theorem}\label{thm:cutoff}
Under Assumption~\ref{ass:shape}, $v_\lambda$ is nondecreasing and right-continuous on $[0,1]$, while $A_\lambda$ and $D_\lambda$ are strictly increasing and right-continuous on $I$. Thus
\[
\Sset_\lambda=[a(\lambda),1],\qquad \piI\leq a(\lambda)<1.
\]
The cutoff is weakly increasing in $\lambda$, and $a(\lambda)\to1$ as $\lambda\to\infty$. There is a unique $\bar\lambda>0$ with $A_{\bar\lambda}(\piI)=0$, and
\[
a(\lambda)=\piI\quad\Longleftrightarrow\quad 0\leq\lambda\leq\bar\lambda.
\]
\end{theorem}
\begin{proof}
Extend every option value by zero below $I$. If $f$ is nonnegative and nondecreasing, kernel order implies that $\M f$ is nondecreasing on $I$. The function $g-\chi_\lambda J_r$ is strictly increasing there. Hence $\mathcal T_\lambda f$ is nondecreasing, including across $\piI$. Iteration from zero proves that $v_\lambda$ is nondecreasing. It follows that $A_\lambda$ is strictly increasing; the increasing transformation in \eqref{eq:D} gives the same property for $D_\lambda$.

The two posterior support maps are continuous. First-order stochastic dominance forces each to be nondecreasing: a fall in the lower support would violate a lower-tail comparison, and a fall in the upper support would violate an upper-tail comparison. Their interior weights are positive. Therefore $\M$ preserves right-continuity of nondecreasing functions. The operator in \eqref{eq:T} preserves that property, including at $\piI$ from the right. Uniform convergence of value iteration gives right-continuity of $v_\lambda$ and hence of both indices. The advice set is an upper interval, contains a neighborhood of one by Proposition~\ref{prop:suppression}, and includes its infimum by right-continuity and the tie rule. This proves its stated form. A possible upward jump at $a$ is allowed; no assertion that $A_\lambda(a)=0$ is needed.

Nesting follows from Theorem~\ref{thm:nesting}. At any fixed $\pi<1$ in $I$, $J_r(\pi)>0$ and \eqref{eq:bound} imply $A_\lambda(\pi)\to-\infty$. Thus the cutoff tends to one.

For the plateau, Corollary~\ref{cor:boundary} gives $A_0(\piI)>0$. The contraction bound implies
\[
\norm{v_{\lambda_2}-v_{\lambda_1}}
\leq\frac{\delta\norm{J_r}}{2(1-\delta)^2}|\lambda_2-\lambda_1|,
\]
where the supremum of $J_r$ can be taken over $I$. Thus $A_\lambda(\piI)$ is continuous in $\lambda$. Nesting and $J_r(\piI)>0$ imply, for $\lambda_2>\lambda_1$,
\[
A_{\lambda_2}(\piI)-A_{\lambda_1}(\piI)
\leq-\frac{\delta(\lambda_2-\lambda_1)}{1-\delta}J_r(\piI)<0.
\]
The value bound also makes this index eventually negative. It crosses zero exactly once, at a positive $\bar\lambda$. Strict increase in the prior and right-continuity then give the plateau assertion.
\end{proof}

\subsection{Sufficient primitive conditions}
\begin{lemma}\label{lem:kernel}
A sufficient condition for posterior-kernel order on an interval is
\begin{equation}\label{eq:kernelorder}
\frac{d e'(\pi)}{[1-e(\pi)q_H][1-e(\pi)q_L]}
\leq\frac1{\pi(1-\pi)}.
\end{equation}
\end{lemma}
\begin{proof}
The upper posterior and its probability $P(\pi)$ are increasing. The derivative of the log odds of the lower posterior is the right side of \eqref{eq:kernelorder} minus its left side, so the lower posterior is nondecreasing. Couple the two-point distributions by a common uniform draw. When the draw switches from lower to upper support, $\phi^-(\pi_1)<\pi_1<\pi_2<\phi^+(\pi_2)$. In all cases the draw is nondecreasing with the prior.
\end{proof}

\begin{proposition}\label{prop:high}
Suppose $c\in C^3[0,1]$ satisfies the model restrictions and $r\in C^3[0,1]$ is increasing and strictly concave, with $r''(1)<0$. There exists $\widehat\pi<1$, independent of $\kappa$, such that a threshold $\piI\geq\widehat\pi$ implies Assumption~\ref{ass:shape}.
\end{proposition}
\begin{proof}
The inverse-function theorem gives $e\in C^2$ and bounded $e'$ near one, and $e_H=e(1)<1$. The left side of \eqref{eq:kernelorder} remains bounded while its right side diverges near one. Kernel order therefore holds on a terminal interval.

The exact variance in Lemma~\ref{lem:effort} and a Taylor expansion of $r$ give, as $x\downarrow0$,
\[
J_r(1-x)=C x^2+O(x^3),\qquad
C=-\frac{r''(1)}2\frac{d^2 e_H}{q_H(1-e_Hq_H)}>0.
\]
The map $J_r$ is $C^2$ near one. Its Taylor coefficients are consequently $J_r(1)=J_r'(1)=0$ and $J_r''(1)=2C$. Taylor expansion of its derivative now gives $J_r'(1-x)=-2Cx+o(x)<0$ near one. This last step uses $C^2$ regularity, rather than differentiating an unspecified remainder. Taking the larger endpoint of the two terminal intervals proves the claim.
\end{proof}

The proposition places implementation in a high-reputation region. It is not a claim that an arbitrary implementation cost generates a cutoff. The restriction on $\kappa$ is $W(p(\widehat\pi))\leq\kappa<W(q_H)$.

\begin{proposition}\label{prop:quadratic}
Let $c(e)=\zeta e^2/2$, $\zeta>q_H$, and $r(\pi)=A+B\pi-\Gamma\pi^2$, where $\Gamma>0$ and $B>2\Gamma$. Put
\[
K_0=\zeta-q_H^2>0,\qquad D_0=q_H d>0,\qquad
\widehat\pi_Q=\frac{D_0-2K_0+\sqrt{D_0^2+4K_0^2}}{2D_0}.
\]
Then $\widehat\pi_Q\in(1/2,1)$, and $\piI\geq\widehat\pi_Q$ is sufficient for Assumption~\ref{ass:shape}. The kernel is ordered on the entire belief interval regardless of this threshold restriction.
\end{proposition}
\begin{proof}
Here $e=p/\zeta$. Condition~\eqref{eq:kernelorder} becomes
\[
d^2\pi(1-\pi)\leq(\zeta-pq_H)(1-pq_L/\zeta).
\]
The left side equals $(p-q_L)(q_H-p)$. Since $\zeta>q_H$ and $p\leq q_H<1$, the two factors on the right are respectively greater than $q_H-p$ and $p-q_L$. Thus the condition holds strictly in the interior. Also,
\begin{equation}\label{eq:quadJ}
J_r(\pi)=\Gamma\frac{d^2\pi^2(1-\pi)^2}{\zeta-p(\pi)^2}.
\end{equation}
Differentiation gives $J_r'\leq0$ exactly when
\[
(2\pi-1)(\zeta-p^2)-p d\pi(1-\pi)\geq0.
\]
For $\pi\geq\piI\geq1/2$, the left side is at least $(2\piI-1)K_0-D_0\piI(1-\piI)$. This quadratic is strictly increasing on $[1/2,1]$, negative at $1/2$, and positive at one; its root is $\widehat\pi_Q$. This proves the claim. Since $W(p)=p^2/(2\zeta)$, the sufficient cost range is $[p(\widehat\pi_Q)^2/(2\zeta),q_H^2/(2\zeta))$.
\end{proof}

\section{Learning under an arbitrary stationary advice set}\label{sec:learning}
Fix a Borel set $\Sset\subseteq I$ and release every favorable certificate on $\Sset$. Let $\pi_0\in\Sset\cap(0,1)$ and
\[
\tau=\inf\{t\geq0:\pi_t\notin\Sset\},\qquad
O(\pi)=\frac\pi{1-\pi},\qquad L(\pi)=\log O(\pi).
\]
Once $\tau$ occurs, beliefs are constant. Denote probability and expectation conditional on ability by $\Pp_\theta,\E_\theta$. Define
\begin{align}
\ell_*&=\frac{1-e_Hq_H}{1-e_Hq_L}>0,\label{eq:ellstar}\\
m_\theta&=\frac12\min_{e\in[e(\piI),e_H]}
\KL\!\left(\operatorname{Ber}(eq_\theta)\,\Vert\,
\operatorname{Ber}(eq_{\theta'})\right)>0,
\quad \theta'\ne\theta.\label{eq:drift}
\end{align}
The minima are positive by continuity, $e(\piI)>0$, and $q_H\ne q_L$.

Here permanent false distrust refers to cessation with a posterior strictly below one when ability is high. It is a failure of eventual vindication, rather than a binary ability classification. In particular, the receiver may still be willing to implement a certificate that the expert withholds.

\Needspace{8\baselineskip}
\begin{theorem}\label{thm:learning}
For every such stationary advice set:
\begin{enumerate}
\item Under low ability, cessation occurs almost surely and
\begin{equation}\label{eq:lowtime}
\E_L\tau\leq
\frac{L(\pi_0)-L(\piI)-\log\ell_*}{m_L}<\infty.
\end{equation}
The posterior at cessation is strictly between zero and one.
\item Under high ability, $\pi_t\to1$ on $\{\tau=\infty\}$, and
\begin{equation}\label{eq:identity}
\Pp_H(\tau<\infty)=\E_L\left[\frac{O(\pi_\tau)}{O(\pi_0)}\right]
\geq\ell_*\frac{O(\piI)}{O(\pi_0)}>0.
\end{equation}
This probability can equal one.
\item Suppose $[b,1]\subseteq\Sset$ for some $b\in(\piI,1)$. High ability has positive probability of continued testing forever if and only if a finite feasible sequence of implemented outcomes can reach $[b,1)$ from $\pi_0$, with every preceding belief in $\Sset$. An empty sequence is allowed. At every initial belief,
\begin{equation}\label{eq:generalupper}
\Pp_H(\tau<\infty)<\frac{O(b)}{O(\pi_0)}.
\end{equation}
When $\pi_0<b$, this upper bound may exceed one and does not itself prove positive survival.
\end{enumerate}
\end{theorem}
\begin{proof}
Before cessation, a favorable certificate occurs with probability $1/2$ under either ability. After it, success multiplies odds by $q_H/q_L$ and failure by $(1-eq_H)/(1-eq_L)$. Otherwise odds are unchanged. Log-odds increments are uniformly bounded. Their conditional means are at most $-m_L$ under $L$ and at least $m_H$ under $H$.

Before stopping, $\pi\geq\piI$. A stopping failure multiplies odds by at least $\ell_*$, while a stopping success raises odds. Consequently,
\[
L(\pi_{\tau\wedge n})\geq L(\piI)+\log\ell_*.
\]
The stopped drift identity gives
\[
\E_L L(\pi_{\tau\wedge n})
\leq L(\pi_0)-m_L\E_L(\tau\wedge n).
\]
Combining these bounds and taking $n\uparrow\infty$ proves \eqref{eq:lowtime}. Every finite update from an interior belief stays interior.

Subtract the conditional drift from the stopped log-odds increments. The resulting martingale differences are bounded. Their average tends to zero almost surely: for example the sum of their variances divided by $t^2$ is finite, and the martingale convergence argument followed by summation by parts gives this strong law. On $\{\tau=\infty\}$, the high-ability drift accumulates at least $m_H$ per period, so $L(\pi_t)\to\infty$ and $\pi_t\to1$.

For each finite public history, its likelihood ratio under $H$ against $L$ is $O(\pi_t)/O(\pi_0)$. Apply this identity to $\{\tau=t\}$ and sum over $t$. Nonnegative summation and $\Pp_L(\tau<\infty)=1$ give \eqref{eq:identity}; no optional-stopping uniform-integrability assertion is required. The lower terminal-odds bound gives its inequality.

If $[b,1]\subseteq\Sset$, every stopping belief is below $b$. The same identity gives the strict inequality \eqref{eq:generalupper}. In particular, from any $x\in[b,1)$ there is positive probability of never stopping. Every finite feasible implemented-outcome sequence has positive probability under $H$. Thus reaching such an $x$ along active predecessors is sufficient for positive survival. Conversely, survival implies convergence to one and therefore reaching $[b,1)$ in finite time. The finite public histories form a countable set after redundant idle periods are removed. A positive-probability hitting event therefore contains a positive-probability finite outcome sequence of the stated kind. This proves necessity.
\end{proof}

Every finite-$\lambda$ equilibrium above contains an upper interval: continuity of $g,J_r$ and $J_r(1)=0$ make $g-\chi_\lambda J_r>0$ near one. Nevertheless, a lower active component need not be able to reach that interval before cessation. Section~\ref{sec:noninterval} supplies an optimal-policy example.

\begin{corollary}\label{cor:cutofflearning}
Under Theorem~\ref{thm:cutoff}, take $\pi_0\in[a,1)$. Cessation can occur only after a failure, and
\begin{equation}\label{eq:cutoffbounds}
\ell_*\frac{O(a)}{O(\pi_0)}
\leq\Pp_H(\tau<\infty)
<\frac{O(a)}{O(\pi_0)}\leq1.
\end{equation}
Hence both permanent false distrust and eventual vindication have positive probability under $H$. Under $L$, the bound in \eqref{eq:lowtime} can use $a$ in place of $\piI$ and the corresponding minimum information rate on $[e(a),e_H]$.
\end{corollary}
\begin{proof}
A success raises the belief and cannot leave $[a,1]$. The stopping failure starts at or above $a$, so $\ell_*O(a)\leq O(\pi_\tau)<O(a)$. Substitute this into \eqref{eq:identity}. The same replacement in the drift proof gives the last claim.
\end{proof}

These stopped-learning conclusions are not specific to optimal effort. A fixed informative effort bounded away from zero gives the same arguments, with its own outcome kernel and likelihood-ratio bounds.

\section{Disconnected advice and a barrier to vindication}\label{sec:noninterval}
We construct a withholding interval between two active regions. The construction first uses a piecewise affine concave payoff and then smooths it while preserving the strict policy inequalities. The resulting income utility is smooth and strictly concave.

\begin{theorem}\label{thm:noninterval}
There is a nonempty open family of primitive parameters with quadratic effort cost and smooth increasing strictly concave career payoff such that:
\begin{enumerate}
\item the posterior kernel is increasing in first-order stochastic dominance, but the optimal advice set is not an interval;
\item a strictly optimal implemented recommendation can end in success and move the expert into strict permanent suppression;
\item from a nonempty lower active region, testing ceases in finite expected time under both abilities, so a high-ability expert is denied eventual vindication with probability one;
\item from an upper active interval, high ability has a positive probability of continued testing and convergence of reputation to one.
\end{enumerate}
The payoff can be generated by the competitive routine-work foundation of Section~\ref{sec:routine}. Failure of a decreasing Jensen cost, rather than failure of kernel order or nonsmooth preferences, is sufficient for these phenomena.
\end{theorem}

\begin{proof}
\textit{Step 1: technology and a barrier interval.}
Choose any $1/2<q_L<q_H<1$, $\zeta>q_H$, and
\[
c(e)=\zeta e^2/2,\qquad
\frac{q_L^2}{2\zeta}<\kappa<\frac{q_H^2}{2\zeta}.
\]
Then $e=p/\zeta$, and Proposition~\ref{prop:quadratic}'s kernel argument gives strictly increasing posterior maps and an ordered kernel everywhere. Choose
\begin{equation}\label{eq:constructak}
\phi^+(\piI)<k<1,\qquad
a=\frac{kq_L}{q_H-dk}.
\end{equation}
Thus $\piI<a<k$ and $\phi^+(a)=k$. Choose $u\in(k,1)$ close enough to $k$ that $\phi^-(u)<k$. Next choose $\ell\in(a,k)$ close enough to $a$ that $\phi^+(\ell)<u$. These choices are possible by continuity and $\phi^-(k)<k=\phi^+(a)$. We have
\begin{equation}\label{eq:barriergeometry}
\piI<a<\ell<k<u<1,\qquad
\phi^+(\ell)<u,\qquad \phi^-(u)<k.
\end{equation}
Finally choose $h>u$ sufficiently near one that $\phi^-(h)>k$.

\textit{Step 2: an auxiliary concave payoff.}
For constants $B>\gamma>0$, let
\[
r^0(\pi)=B\pi-\gamma(\pi-k)_+.
\]
It is increasing and concave. At $a$, both possible posteriors are at most $k$; at $h$, both are above $k$. The payoff is affine on each of those posterior supports, so $J_0(a)=J_0(h)=0$. Thus $A_\lambda(a)\geq g(a)>0$ and $A_\lambda(h)\geq g(h)>0$ for every finite $\lambda$.

Throughout $[\ell,u]$, the posterior support strictly straddles $k$. Bayes' rule and cancellation of the affine part give
\begin{equation}\label{eq:hingeJ}
J_0(\pi)=
\begin{cases}
\gamma P(\pi)[\phi^+(\pi)-k],&\pi\leq k,\\
\gamma[1-P(\pi)][k-\phi^-(\pi)],&\pi\geq k.
\end{cases}
\end{equation}
The two expressions agree at $k$. Since $P$ and both posterior maps are increasing,
\begin{equation}\label{eq:jstar}
J_0(\pi)\geq j_*:=\gamma\min\left\{
P(\ell)[\phi^+(\ell)-k],\,
[1-P(u)][k-\phi^-(u)]\right\}>0
\end{equation}
on the entire barrier interval. Fix any $\delta\in(0,1)$ and choose
\begin{equation}\label{eq:constructlambda}
\lambda>\frac{G(2-\delta)}{2\delta j_*},\qquad
m:=\chi_\lambda j_*-\frac{G(2-\delta)}{2(1-\delta)}>0.
\end{equation}
The general value bound implies $A_\lambda\leq-m$ on $[\ell,u]$. This is a strict statement about the exact Bellman fixed point, not a myopic approximation.

\textit{Step 3: smooth strictly concave utility.}
For $\eta>0$, define
\begin{equation}\label{eq:smoothpayoff}
H_\eta(x)=\frac{x+\sqrt{x^2+\eta^2}}2,
\qquad
r^\eta(\pi)=B\pi-\gamma[H_\eta(\pi-k)-H_\eta(-k)].
\end{equation}
This payoff is $C^\infty$, vanishes at zero, and satisfies
\[
(r^\eta)'=B-\gamma H_\eta'>B-\gamma>0,
\qquad (r^\eta)''=-\gamma H_\eta''<0.
\]
The inequality $0\leq H_\eta(x)-x_+\leq\eta/2$ and cancellation of constants in $J$ imply
\begin{equation}\label{eq:smoothingbound}
\sup_{\pi\in[0,1]}|J_\eta(\pi)-J_0(\pi)|\leq\gamma\eta.
\end{equation}
Choose
\begin{equation}\label{eq:etachoice}
0<\eta<\frac{\min\{g(a),g(h),m\}}{\chi_\lambda\gamma}.
\end{equation}
Strict concavity gives $J_\eta\geq0$, so the same general value bound applies. At $a$ and $h$,
\[
A^\eta_\lambda(a)\geq g(a)-\chi_\lambda\gamma\eta>0,
\qquad A^\eta_\lambda(h)\geq g(h)-\chi_\lambda\gamma\eta>0.
\]
For every $\pi\in[\ell,u]$,
\[
A^\eta_\lambda(\pi)\leq-m+\chi_\lambda\gamma\eta<0.
\]
Hence two strictly active regions are separated by a strictly inactive interval. Strictness and continuity of the displayed lower bounds also give an active neighborhood of $a$ below $\ell$. A success at $a$ leads exactly to $k\in[\ell,u]$ and therefore to permanent suppression.

\textit{Step 4: the barrier prevents vindication from below.}
Start at any active $\pi_0<\ell$. Every subsequent active belief stays below $\ell$: failure lowers it, while success leads to at most $\phi^+(\ell)<u$ and cannot cross the inactive interval. At cessation the belief is also below $u$. Thus $L(\pi_{\tau\wedge n})<L(u)$ for every $n$. Under high ability its stopped drift is at least $m_H$, so
\[
\E_H(\tau\wedge n)\leq\frac{L(u)-L(\pi_0)}{m_H}.
\]
Letting $n\uparrow\infty$ proves finite expected cessation under $H$. Theorem~\ref{thm:learning} already proves it under $L$. At a finite stopping date the posterior remains interior, and no further evidence arrives. Conversely, advice is strictly active near one, and the reachability result gives positive high-ability survival from that upper region.

\textit{Step 5: primitives and robustness.}
For any wage interval with $\Delta w>0$, set $U(w)=r^\eta((w-w_L)/\Delta w)$ on that interval, up to an additive constant. It is smooth, increasing, and strictly concave, and \eqref{eq:wage} generates exactly $r^\eta$. All parameter and sign restrictions above are strict and depend continuously on the finite primitive parameters. For nearby parameters, define $a$ by \eqref{eq:constructak} and retain the strict barrier inequalities. Thus the construction describes a nonempty open family. The auxiliary kink is only a proof device; the stated model uses $r^\eta$ with $\eta>0$.
\end{proof}

The economic source of the gap is concentrated career risk. Near the middle wage range, an experiment crosses a region where marginal income utility falls sharply. At lower beliefs both possible wages are approximately on one side of that range; near certainty they are on the other. An informative success can therefore move the expert into a region where further evaluation is too costly. This interpretation is consistent with increasing effort and an ordered posterior kernel.

The construction does not claim that disconnected advice is typical, or that its logic requires optimized effort. It proves why the shape restriction in the cutoff theorem has substantive content. It also shows that positive vindication probability from every active state must not be carried over to arbitrary advice sets.

\section{Outside reports: proofs}\label{app:reports}
\subsection{Proof of Proposition~\ref{prop:reports}}
\begin{proof}
The Bellman operators contract by $\delta$, and their values lie between zero and $C_0/(1-\delta)$. Thus $\norm{(\Q_\varepsilon-\Id)V_0}\leq\varepsilon C_0/(1-\delta)$. Comparing the two operators at $V_0$ and using contraction proves \eqref{eq:reportsVbound}. It follows that
\[
\norm{\Q_\varepsilon V_\varepsilon-V_0}
\leq\frac{\varepsilon C_0}{(1-\delta)^2}.
\]
Since $\norm{(\M-\Id)f}\leq2\norm f$, equation~\eqref{eq:reportsD} yields \eqref{eq:reportsDbound}. A margin of magnitude $m>0$ survives whenever $2\delta\varepsilon C_0/(1-\delta)^2<m$.

For positive reporting rates, infinitely many routine outputs are revealed almost surely. Their success frequency converges to $a_\theta$, which differs across abilities. Ability is therefore measurable with respect to the infinite public record. The bounded posterior martingale converges to its conditional expectation on that record, namely $\mathbf1_{\{\theta=H\}}$. Adaptive project observations do not remove the identifying routine observations.

Before the first public routine report, the initial strict withholding belief does not change and no project is tested. Let $T$ count periods through that first report; it is geometric with mean $1/\varepsilon$. Because reports arrive after the project decision, $T_{\rm test}\geq T$, proving \eqref{eq:wait}. These are lower bounds, not an exact recovery-time formula. The baseline index has a positive uniform margin on a sufficiently short upper interval. For sufficiently small $\varepsilon$, that margin survives. Under $H$, convergence to one therefore puts the process permanently in this active upper interval, and favorable assessments arrive infinitely often. Project testing resumes almost surely.
\end{proof}
\subsection{Proof of Proposition~\ref{cost:prop:reports}}
\begin{proof}
Let $D_i^\varepsilon$ denote the unnormalized release-versus-silence gain. By Proposition~\ref{prop:reports}, the Bellman perturbation bound is
\[
 \|D_i^\varepsilon-D_i^0\|_\infty
 \leq\frac{2\delta\varepsilon C_i}{(1-\delta)^2},\qquad
 C_i=\beta\max r_\rho+\frac{\alpha G_i}{2}.
\]
At $\pi_0$, $D_1^0=\alpha A_1<0$ and
$D_2^0=\alpha\,2(1-\delta)A_2/(2-\delta)>0$.
Thus choosing the displayed error smaller than both strict margins preserves the initial actions. In economy 1, without a report the state remains $\pi_0$ and the expert stays silent. If $T\in\{1,2,\ldots\}$ is the earliest period in which a report arriving after a previous project can affect implementation, then $T$ is geometric with parameter $\varepsilon$. No project occurs before $T$ in economy 1. At any later period, expected client surplus is at most $G_1/2$, since a favorable certificate has probability $1/2$. Consequently
\[
 0\leq \mathcal W_1^\varepsilon(\pi_0)
 \leq\frac{G_1}{2(1-\delta)}
 \frac{\varepsilon\delta}{1-(1-\varepsilon)\delta}.
\]
In economy 2, the initial project gives
$\mathcal W_2^\varepsilon(\pi_0)\geq g_2(\pi_0)/2$.
A simple additional sufficient restriction is
\[
 \varepsilon<\frac{g_2(\pi_0)(1-\delta)^2}{\delta G_1}.
\]
It makes the first upper bound strictly smaller than the second lower bound. All requirements admit positive $\varepsilon$.
Finally, independent informative routine reports arrive infinitely often almost surely. Their likelihood contributions identify ability; other Bayesian public evidence does not erase that identification (Proposition~\ref{prop:reports}). Thus both economies learn the true type in the limit.
\end{proof}

\Needspace{8\baselineskip}

\end{document}